# Galactic Cosmic Rays and Insolation are the Main Drivers of Global Climate of the Earth

V.D. Rusov[1], I.V. Radin[1], A.V. Glushkov[1], V.N. Vaschenko [2], V.N. Pavlovich [3], T.N. Zelentsova[1], O.T. Mihalys[1], V.A.Tarasov[1], A. Kolos [1]

[1] *Odessa National Polytechnic University, Shevchenko av. 1, 65044 Odessa, Ukraine*
[2] *Ukrainian National Antarctic Center, Peremogi av.10, 01033 Kiev, Ukraine*
[3] *Institute for Nuclear Research, Nauki av. 47, 03028 Kiev, Ukraine*

**Abstract**

An energy-balance model of global climate, which takes into account a nontrivial role of galactic cosmic rays, is developed. The model is described by the fold catastrophe equation relative to increment of temperature, where galactic cosmic rays and insolation are control parameters. The comparison of the results of a computer simulation of time-dependent solution of the presented model and oxygen isotope records of deep-sea core V28-238 over the past 730 kyr are presented. The climate evolution in future 100 kyr is also predicted.

---

[*] Corresponding author: Prof. Rusov V.D., The Head of Department of Theoretical and Experimental Nuclear Physics, Odessa National Polytechnic University, Shevchenko av. 1, Odessa, 65044, Ukraine; Phone/Fax: + 380 482 641 672, E-mail: siiis@te.net.ua

## 1. Introduction

It is well known that galactic cosmic rays (GCR's) play one of the key roles in the mechanism of weather and climate formation on our planet [1, 2]. Results of numerous investigations of the influence of GCR fluxes on the atmospheric processes, in particular, on charged aerosols formation, which are the centers of water vapour condensation (the basic greenhouse gas), indicate the following causal sequence of events [1-3]: brighter sun $\Rightarrow$ insolation and solar activity changes $\Rightarrow$ modulation of GCR flux $\Rightarrow$ cloud coverage and thunderstorm activity changes $\Rightarrow$ albedo changes $\Rightarrow$ weather and climate changes.

Until now, there is no clear understanding, whether this causal sequence of GCR flux modulation characterizes only the weather or only the climate. We think that the answer to this question is contained in the known spectrum of air temperature variations in the North-Atlantic sector of the terrestrial globe [4]. It is evident from the frequency spectrum of air temperature variations (Fig.1), the part of weather changes - characterized by the white noise - is situated right side (> 0.1 cycle/year). On the left side (< $10^{-3}$ cycle/year) there is a spectrum of climate long-periodic variations characterized by so-called "red" noise. It, unlike the white noise, possesses a certain degree of predictability.

Thus, analysis of the temperature variation spectrum of Earth's climate system (ECS) [4] shows that time averaging of ECS parameters is required for the correct statistical description of weather or climate. It is obvious that the time scale of averaging of ECS parameters should be less than 20 years and more than 1000 years for stochastic description of the weather and for deterministic description of the climate evolution, respectively.

Hence, GCR flux modulation by solar wind in the above mentioned causal sequence of events determines changes of the weather, because this modulation takes place on the time intervals from a few days (Forbush phenomenon) up to ten years (11-year solar cycle). Therefore such a type of GCR flux modulation is not related to the climate and even more can not be used as a control parameter in the models of global climate of the Earth.

On the other hand, we can consider the spectral density of virtual axial dipole moment (VADM) variations of terrestrial magnetic field for the past 800 kyr [5] shown in Fig. 2. The presence of characteristic short and intermediate periods of magnetizing force variations can be observed in the order of ~100 kyr and ~ 410 kyr, respectively. They coincide with eccentricity variations of the elliptic orbit of the Earth. Possible explanation of this effect - on phenomenological level - can be based on the supposition that the gravitation field of the Sun acting upon convection (hydromagnetic dynamo) in the liquid core of the Earth "delegates" the intensity variations to terrestrial magnetic field with the periods corresponding to the periods of eccentricity variations.

Thus, spite the fact that the mechanism of appearance of the frequencies typical for eccentricity in the spectrum of magnetizing force variations is not quite clear it is possible to conclude that only the secular variations of terrestrial magnetic field intensity are specific for GCR flux modulations on the time scale larger than 1000 years. It is obvious that such type of GCR flux modulation has direct relation to the climate, and it can be used as one of the control parameters in the models of the global climate of the Earth.

The purpose of present paper was a construction of the energy-balance model of climatic response to orbital variations, which takes into account influence of cosmic rays on the global climate of the Earth.

**2. Theory**

By virtue of the law of conservation of energy the real thermal radiating power of the Earth is the difference between the self-radiation power of the Earth's warming surface $I(T, t)$ and thermal energy power $G(T, t)$ stored by the greenhouse gas (only water vapour gas is considered for the sake of simplicity). Because the radiant equilibrium is reached in times in the order of several tens of thousand years, consideration of the greenhouse effect makes it possible to write down the following energy-balance equation for ECS:

$$U(T,t) = P(t)[1-\alpha(T)] - I(T) + G(T,t), \qquad (1)$$

where the left side of the equation (if it is not equal to zero) describes the value of so-called "inertial" power of heat variations in ECS; the term $P(t) = (1/4)S(t)\gamma$ is the heat flow of solar radiation to the upper boundary of the atmosphere, W; $S$ is insolation; $\gamma$ is the area of the external boundary of the upper atmosphere, m$^2$; $\alpha$ is ECS albedo; $I = \gamma(\sigma T^4)$, W; $T$ is the temperature of ECS, K; $t$ is the time of energy balance consideration.

The temperature dependence of ECS albedo is taken as the following parameterization:

$$\alpha = \alpha_0 - \eta_\alpha (T - 273), \qquad (2)$$

which reflects well the behavior of ECS albedo at $\alpha_0 = 0.315$ and $\eta_\alpha = 0.035$ in the temperature range 262–270 K corresponding to the experimental range of the average air temperatures at "Vostoc" station region for last 420 kyr [6].

Now let us consider the temperature dependence of the rate of heating energy $G(T, t)$ stored by the greenhouse gases. It was experimentally shown that the basic source of air ionization in the tropo- and stratosphere are galactic cosmic rays, and the dependence of ion concentration in the atmosphere $n$ on GCR flux $\Phi$ is linear, i.e. $n \sim \Phi$ [2]. Since both the insolation and GCR flux are modulated by eccentricity *e(t)*, it is possible to suppose that the total energy of water vapour condensation on charged aerosols is proportional to the energy stored by the greenhouse gases *G(T, t)*.

Taking into account that the heat emitted as a result of water vapour condensation generates temperature pulsations in the turbulent air flow with Kolmogorov spectrum (a horizontal part of the cell of global atmosphere circulation), we can use the Obukhov law [7]. According to it the structure of the temperature field in turbulent conditions is determined not only by the dissipation rate of turbulence kinetic energy per mass unit $\varepsilon$, but also by the dissipation rate of temperature fluctuation intensity $N_T$, which is the order of magnitude equal to:

$$N_T \cong (\Delta T)^2 \Delta u L^{-1}, \qquad (3)$$

where $\Delta u$ is the characteristic size of the scale of main energy-carrying vortexes speed; $\Delta T$ is the characteristic temperature difference in flow on its external scale *L*.

Next it is possible to present generalized Kolmogorov-Obukhov law [7], which takes into account both the turbulent pulsations of kinetic energy and the temperature pulsations, in an equivalent spectral (on space) form. Let $E_T(k)dk$ is the kinetic energy (per unit mass of liquid) contained in pulsations with values *k* in given interval *dk*. As the dependence $E_T(k)$ has a dimension of cm$^3$/s$^2$, making the combination of this dimensionality from $N_T$, $\varepsilon$ and *k* in the inertial interval of scales we obtain the following expression $k \in [k_L, \infty]$:

$$E_T(k) = C_{1T} N_T \varepsilon^{-1/3} k^{-5/3}, \qquad . \qquad (4)$$

where $C_{1T} \approx 1.4$

By integration of Eq. (4) in the range $k \in [k_L, \infty]$ and taking into account the order of the magnitude of energy dissipation $\varepsilon \sim (\Delta u)^3/L$ and temperature fluctuation intensity Eq. (3), we obtain the estimate of heat transported by the global circulation to the atmosphere bottom layer:

$$E_T = C_{1T} \frac{N_T}{\varepsilon} \left(\frac{\varepsilon}{k_L}\right)^{2/3} = C_{1T}(\Delta T)^2. \qquad (5)$$

Since transported heat Eq.(5) (for the characteristic temperature difference $(T_{out}-T)$ in the turbulent flow on its external scale *L*) is proportional to the total energy of water vapour condensation and therefore to *G(T, t)*, hence taking into account the GCR flux modulation by eccentricity *e(t)*, it is possible to write down the following expression for *G(T, t)*:

$$G(T,t) = g(\Delta T)^2 = k\Phi(t)e(t)(T_{out} - T)^2, \qquad (6)$$

where $\Phi(t)=N_\Phi(t)\cdot\gamma$, $N_\Phi(t)$ is GCR flux intensity, m$^{-2}$s$^{-1}$; $k$ is a constant, GeV/(mK)$^2$, which in zonal ECS models generally depends on the latitude.

Finally, collecting all partial contributions of heat flows (Eqs. (2) and (6)) and considering $I=\gamma(\sigma T^4)$ to the resulting energy-balance (Eq.(1)), we obtain:

$$-\frac{1}{4\gamma\sigma}U(T,t) = \frac{1}{4}T^4 + \frac{1}{2}a(t)(T_{out} - T)^2 + b(t)T + absolute\ term, \quad (7)$$

$$a(t) = -g/2\gamma\sigma, \quad b(t) = -\eta_\alpha S/16\sigma.$$

It is obvious that Eq. (7) describes the collection of energy-balance functions $U(T, a, b)$, which depend on two control parameters $a(t)$ and $b(t)$. It is not difficult to see that this collection represents so-called "potential" of fold catastrophe [8]. From hereon we will be interested in the fold catastrophe (Eq. (7)) relative to increment $\Delta T = T-T_0$ of $U(T_0+\Delta T, a, b) - U(T_0, a, b) = \Delta U$ type, where $T_0$ is the average temperature (averaged over respective time interval $\Delta t$). Here the increment of the first term of right side of Eq. (7) was used without the cubic item of $\sim (\Delta T)^3$ type due to following approximation:

$$(T_0 + \Delta T)^4 - T_0^4 \cong 7\cdot 10^{-3} T_0^3 (\Delta T)^4 + 4T_0^3 \Delta T, \quad for \quad \Delta T = 0 - 4\ K, \quad (8)$$

where average error does not exceed 0.01% in the given temperature interval.

Let us remind that the normalized variation of insolation $\Delta\hat{S}=(S-S_0)/\sigma_S$ with the average $<\Delta\hat{S}>=0$ and dispersion $\sigma^2_{\Delta\hat{S}}=1$ is applied for simulation of the ECS ($S_0=P(t=0)/\gamma$ is insolation in units of W/m$^2$ at the time of $t=0$.

Constructing Eq.(7) type relative to temperature disturbance $\Delta T$ we thus obtain the following expression for the increment of heating rate $\Delta U$:

$$V(\Delta T, t) = \frac{1}{4}\Delta T^4 + \frac{1}{2}\tilde{a}(t)\Delta T^2 + \tilde{b}(t)\Delta T, \quad V = -\frac{35.71}{\gamma\sigma T_0^3}\Delta U, \quad (9)$$

$$\tilde{a}(t) = -\frac{71.42 k N_\Phi}{\sigma T_0^3} e(t) = -\tilde{a}_0 e(t), \quad (10)$$

$$\tilde{b}(t) = -\frac{8.91\eta_\alpha \sigma_S}{\sigma T_0^3}\left\{\Delta\hat{S}(t) + \frac{1}{\sigma_S}\left[S_0 - \frac{16\sigma T_0^3}{\eta_\alpha}\right]\right\} = -\tilde{b}_0\left[\Delta\hat{S}(t) + \xi_G\right]. \quad (11)$$

The canonical form of the variety of the fold catastrophe, which represents point set $(\Delta T, \tilde{a}, \tilde{b})$, satisfies the equation:

$$\nabla V(\Delta T, t) = \Delta T^3 + \tilde{a}(t)\cdot\Delta T + \tilde{b}(t) = 0. \quad (12)$$

Thus, the general bifurcation problem of the determination of solution $\Delta T(t)$ is reduced to solving Eq. (12) for the appropriate joint trajectory $\widetilde{a}(t), \widetilde{b}(t)$ in the space of control parameters.

## 3. Computer simulation

Now we can determine the solution $\Delta T(t)$ of the bifurcation problem Eq.(12), for example, at latitude 65° N. It is obvious that the basic Eq. (9) and bifurcation Eq. (12) within the framework of any zonal energy-balance climate model formally save their form, and only require the change of control parameters $\widetilde{a}(t), \widetilde{b}(t)$ at latitude 65° N. The userfulness of zonal climatic model is exhibited with an addition of an advection term in Eqs. (1) and (12), which describes the flows of explicit and latent heat through the lateral faces of the element (zone) of ECS in the form of additive constant $A_{65°N}$ in value $\xi_G$ of control parameter $\widetilde{b}(t)$ in Eq. (11).

To solve the bifurcation Eq. (12), which describes the extreme values of the increment of temperature $\Delta T$ for the element (zone) of ECS at latitude 65° N, it is necessary to determine two climatic constants $k\Phi_{65°N}$ and $\xi_G(A_{65°N})$ in Eqs. (10) and (11). Values of remaining parameters (except for the average temperature $T_0$) are known. For example, the values of eccentricity $e(t)$ (Fig. 3a) and normalized insolation $\Delta \hat{S}(t)$ (Fig. 3b) were calculated by Berger [9]. The value of mean-root-square error of insolation variation is equal to $\sigma_S = 21$ for the selected time period including 730 kyr in the past and 100 kyr in the future.

Selection of ECS average temperature $T_0$ at latitude 65°N was done through following considerations. It is well known that the value of modern climatic representative temperature is about 268.6 K at latitude 65°N. On the other hand, according to the data of Russian-French-Italian researches from an ice core from borehole at Russian Antarctic station "Vostoc" [6] the range of average increment $\Delta T = [2, -6]$ is less approximately for 2 K compare to the modern temperature. Therefore the average temperature of $T_0 = 266.6$ K was used for further calculations.

The traditional calibration method relative to experimental data was applied for the calculation of climatic constants $k$ and $\xi_G(A_{65°N})$. The essence of this method lies as follows. According to the experiments, which were made at Antarctic station "Vostoc" [6], it is possible to conclude that the jump of the temperature $\Delta T$ relative to the average temperature $T_0 = 266.6$ K was approximately $\Delta T \cong 4$ K at $t = 120$ kyr ago. Such a supposition results in the following single-valued form of bifurcation equation (12): $(\Delta T)^3 - 12\Delta T - 16 = 0$ at $t = 120$ kyr.

Hence, the fixed values of control parameters $a_{120} = 12$ K$^2$ and $b_{120} = 16$ K$^3$ at $t = 120$ kyr in the past at latitude 65° N make it possible to determine the values of climatic constants $k\Phi_{65°N}$ and $\xi_G(A_{65°N})$ from Eqs. (10) and (11). Taking into account that at $t = 120$ kyr in the past to eccentricity $e_{120} = 0.038$ there corresponds the value of the normalized variation of insolation $\Delta \hat{S}_{120} = 2.3$, and also using the standard error of insolation variations $\sigma_S = 21$, $\eta_\alpha = 0.035$ K$^{-1}$ and $\sigma = 5.67 \cdot 10^{-8}$ W/m$^2$K$^4$, we obtain the values of two climatic constants, i.e. $k\Phi_{65°N}/\gamma = 4.42$ and $\xi_G(A_{65°N}) = 0.25$ at the average temperature of $T_0 = 266.6$ K:

The time-dependent solution of bifurcation Eq. (12) obtained with computer simulation, which describes the temporal changes of the increment of temperature $\Delta T$ relative to the average temperature $T_0 = 266.6$ K for 730 kyr in the past and 100 kyr in the future, is shown in Fig. 3d. In Fig. 3c oxygen isotope curve for deep-sea core V28-238 from the Pacific Ocean over the past 730 kyr is presented [10]. Data from Ref. [10] are plotted against the PDB standard on the time scale of Kominz et al. [10]. Good agreement between experimental (Fig. 3c) and theoretical (Fig. 3d) data is the peculiar indicator of high quality of prognosis of temporal changes of global temperature $T_0 + \Delta T$ at latitude 65°N over the next 100 kyr (Fig. 3d).

## 4. Conclusions

Thus, the main statement of the proposed model is that the global climate of the Earth is completely determined by two control parameters - galactic cosmic rays and insolation - and practically has not limitations on the horizon of global forecast, i.e. is quite predictable.

**FIGURE CAPTIONS**

Fig. 1. Combined spectrum of the air temperature variations in the North-Atlantic sector of the terrestrial globe [4]: $f$ are frequencies, cycle/year; $S(f)$ is spectral density; 1 – Central England, paleobotany; 2 – the same, chronicles; 3 – Iceland, chronicles; 4 – Greenland, obtained by $\delta^{18}O$; 5 – Central England, by Manley series [4].

Fig. 2. Spectrum of VADM synthetic curve variations over the past 800 kyr [5].

Fig. 3. Model of climatic response to orbital and insolation variations compared with isotopic data on climate of the past 730 kyr. Variations in orbital eccentricity (a) and insolation (b) at 65°N at the summer solstice over the past 730 kyr and over the future 100 kyr [9]; (c) oxygen isotope curve V28-328 [10] from the Pacific Ocean (PDB standard) on the time scale by Kominz et. al. [10]; (d) result of our model calculated by Eq. (18): evolution of the increment of temperature $\Delta T$ relative to the average temperature $T_0$=266.6 K over the past 730 kyr and 100 kyr in future.

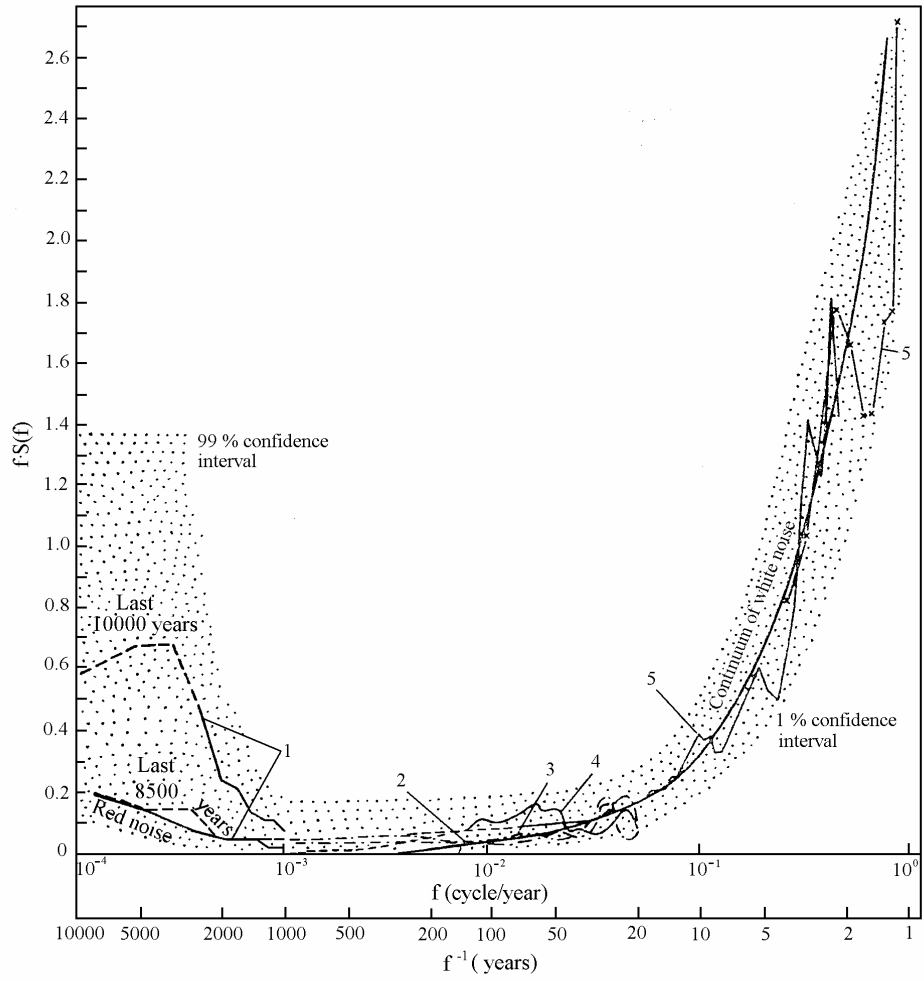

Fig.1.

Rusov V.D. et al.

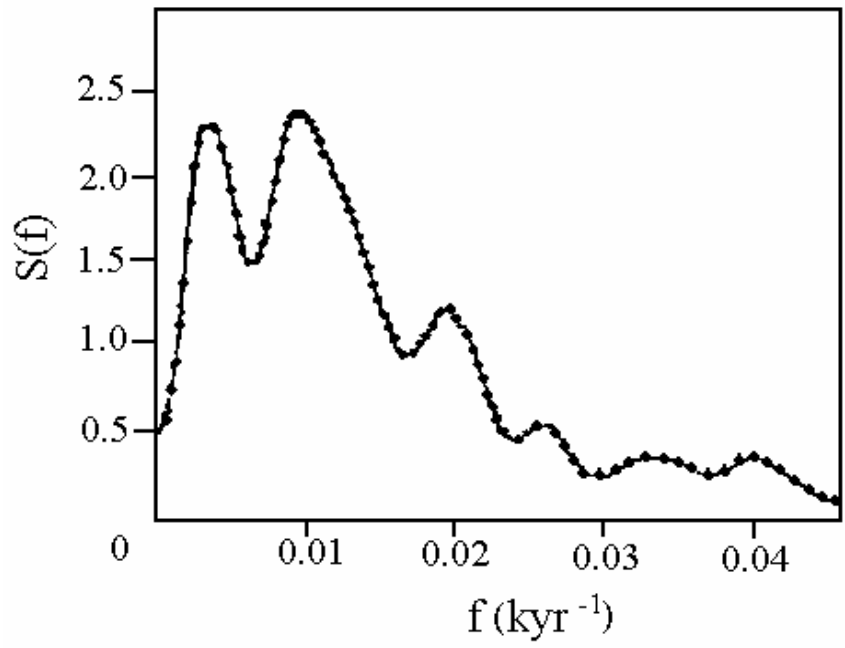

Fig.2.

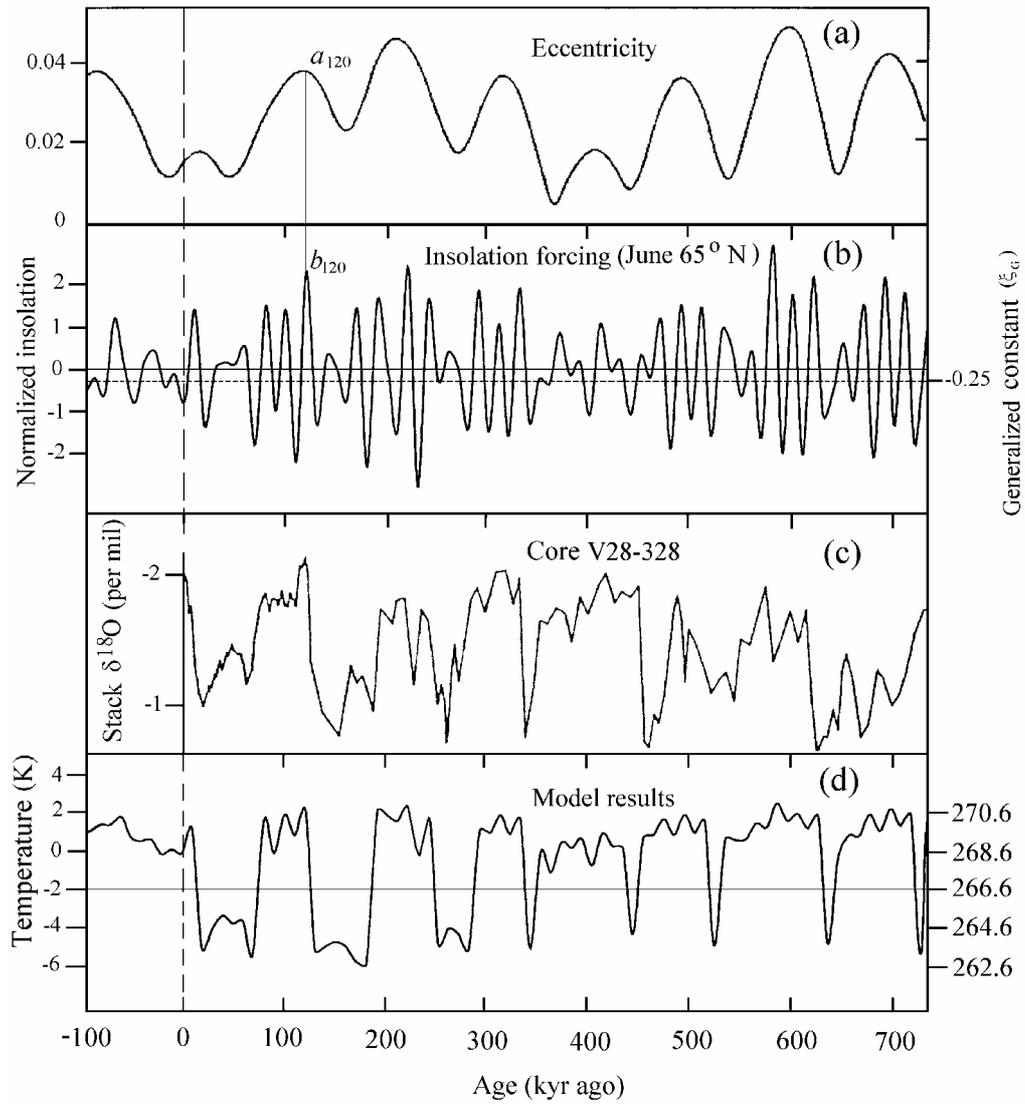

Fig.3.

Rusov V.D. et al.